\documentclass[12pt]{article}
\usepackage{amsmath,amssymb,mathrsfs,url,graphicx,color}
\usepackage{cite}

\title{Quantum Einstein equations}

\author{Detlef D\"urr\footnote{Mathematisches Institut, Ludwig-Maximilians-Universitat M\"{u}nchen, Theresienstr.\ 39, D-80333 M\"{u}nchen, Germany} and Ward Struyve\footnote{{Instituut voor Theoretische Fysica, KU Leuven, Belgium}} \footnote{Centrum voor Logica en Filosofie van de Wetenschappen, KU Leuven, Belgium}  }

\def\de{\delta}

\def\ka{\kappa}

\def\pa{\partial}

\def\al{\alpha}

\def\ka{\kappa}
\def\ii{\textrm i}
\def\ee{\textrm e}

\newcommand{\be}{\begin{equation}}
\newcommand{\en}{\end{equation}}
\newcommand{\bi}{\begin{itemize}}
\newcommand{\ei}{\end{itemize}}

\begin{document}
\maketitle

\begin{abstract}
  \noindent
We derive the quantum Einstein equations (which are the quantum generalisation of the Einstein equations of classical gravity) from Bohmian quantum gravity. Bohmian quantum gravity is a non-classical geometrodynamics (in the ADM formalism) which describes the time evolution of a 3-geometry and of a matter field (or other matter degrees of freedom) on a three manifold. The evolution is determined by a velocity law which is defined by the wave function. The wave function itself satisfies the Wheeler-DeWitt equation. We cast the Bohmian dynamics into the form of the Einstein field equations, where the interesting novelty is a contribution to the energy-momentum tensor that depends on the quantum potential.
\end{abstract}

\section{Introduction}
Non-relativistic Bohmian mechanics \cite{bohm93,holland93b,duerr09,duerr12} describes the motion of point-particles whose velocity depends on the wave function. For a single particle moving on a 3-dimensional Riemannian manifold with metric $g_{ij}(x)$, $x = (x^1,x^2,x^3) $, the wave function $\psi(x,t)$ satisfies the Schr\"odinger equation (in units such that $c=\hbar=1$)\footnote{The generalization to many particles is straightforward.}
\be
\ii \frac{\pa \psi}{\pa t} = -\frac{1}{2m} \nabla^2\psi + V\psi ,
\en
where $\nabla^2$ is the Laplace-Beltrami operator with respect to the metric $g_{ij}$, i.e.,
\be
\nabla^2 \psi = g^{ij} \nabla_i \nabla_j \psi  = \frac{1}{\sqrt g} \pa_i {\sqrt g} g^{ij}\pa_j \psi.
\en
The equation of motion for the point-particle (the {\it guidance equation}) with position $x(t)$ is
\be
\dot  x= \frac{1}{m} \nabla S ,
\label{1}
\en
where $\psi=|\psi|\ee^{\ii S}$, or in component form:
\be
\dot x^i= \frac{1}{m}\nabla^i S = \frac{1}{m} g^{ij}\frac{\pa S}{\pa x^j}.
\label{1b}
\en

By differentiating \eqref{1} with respect to $t$, one finds the Newtonian like equation
\be
m \ddot  x = -  \nabla (V + Q),
\label{2}
\en
where 
\be
Q = - \frac{1}{2m} \frac{\nabla^2|\psi|}{|\psi|}
\en
is the {\em quantum potential}. The quantum force $ - \nabla Q$ may be seen as a ``quantum signature'' signifying the difference between Bohmian motion and classical motion (i.e., Newtonian motion in the potential $V$).

The Newtonian equations \eqref{2} are derivable from the action
\be
{\mathcal S} = \int dt \left[  \frac{m}{2} g_{ij }(x)\dot x^i \dot x^j  - V(x) - Q(x,t)\right].
\label{3}
\en
This action can also be expressed in terms of phase space variables:
\be
{\mathcal S} = \int dt \left[   p_i \dot x^i   - H(x,p,t)\right],
\en
where
\be
H(x,p,t) = \frac{1}{2m} g^{ij}(x)p_ip_j + V(x) + Q(x,t).
\en
The corresponding equations of motion are then Hamilton's equations:
\be
\dot  x^i =  \frac{1}{m} g^{ij}p_j  ,\qquad \dot  p_i = - \pa_i (V+Q). 
\label{9}
\en
These equations of course do not entail \eqref{1}. Equation \eqref{1} must  be imposed separately. This can be achieved by setting $p_i= \pa_i S$ initially. Then $p_i= \pa_i S$ at all times, and \eqref{1} is satisfied. In other words, the Bohmian dynamics can be derived from an action principle leading to a Hamiltonian description which must however be subjected to special initial conditions. (It is interesting to note that the guidance equation corresponds to the classical Hamilton equation $\dot  x^i =  \frac{1}{m} g^{ij}p_j$ with the replacement $p_i= \pa_i S$. This results in some formal analogy with the classical Hamilton-Jacobi theory which is also present in the case of Bohmian quantum gravity.)

In this paper, we will consider Bohmian quantum gravity for a scalar matter field \cite{horiguchi94,shtanov96,goldstein04,pinto-neto05a,pinto-neto19}. Bohmian quantum gravity rids orthodox quantum gravity from the measurement problem as well as the problem of time. It describes the evolution of a 3-metric (in the ADM formulation of geometrodynamics) whose velocity depends on the wave function, which satisfies the Wheeler-DeWitt equation. The guidance equation is a first order equation, similar to \eqref{1}. Even though the wave function is static, the actual 3-metric generically is not, which solves the problem of time. The goal is to formulate the second order dynamics, which can be cast in the form of the Einstein field equations. The resulting equations, which we will call the {\em quantum Einstein equations}, will depend on a quantum potential. Two different ways will be presented to find these equations. The first one is by using an action for the Bohmian second order dynamics (similarly as with \eqref{3} in non-relativistic Bohmian mechanics). The second one proceeds more directly by considering what the Bohmian equations of motion entail for the dynamics of the Einstein tensor.   

The quantum Einstein equations were first considered in \cite{shojai98,shojai02,shojai04}, and although the strategy in there seems to be along the lines as presented here, the reported results are not correct as shall be indicated later on. An immediate consequence of the quantum Einstein equations is that the total Bohmian energy-momentum tensor is covariantly conserved. As shall be explained in section \ref{energy}, this should be contrasted to the situation for other Bohmian theories where the classical energy conservation law does not apply. Furthermore, the quantum Einstein equations allow for a study of the classical limit and possible deviations of classicality (which are important in for example the big bang era \cite{pinto-neto13,falciano15,pinto-neto19} and possibly beyond \cite{demaerel20}). It may also be the starting point for the development of a semi-classical theory of gravity based on Bohmian mechanics \cite{struyve15,struyve17a}. 

The outline of the paper is as follows, we will first revisit the classical Einstein-Hilbert action in terms of phase-space variables in section \ref{classical} and Bohmian quantum gravity in section \ref{bqm}. In sections \ref{quantumeinstein1} and \ref{quantumeinstein2}, we will present the two different ways to derive the quantum Einstein equations. In section \ref{Mini-superspace}, the quantum Einstein equations are presented for the case of mini-superspace. In section \ref{energy}, we comment on the issue of energy-momentum conservation. We conclude in section \ref{con}.

\section{Classical gravity and geometrodynamics}
\label{classical}
The action for general relativity coupled to a scalar field is given by (see e.g.\ \cite{carlip19} whose sign convention $(+,-,-,-)$ is adopted)
\be
{\mathcal S}_{\textrm{cl}} = \int d^4x  {\mathcal L}_{\textrm{cl}}={\mathcal S}_G + {\mathcal S}_M = -\frac{1}{\ka} \int d^4 x   \sqrt{-g} (R + 2 \Lambda) + \int d^4 x  \sqrt{-g} \left( \frac{1}{2} g^{\mu \nu} \pa_\mu\phi \pa_\nu\phi - {\mathcal U}(\phi) \right),
\label{10}
\en
which is the sum of the free gravitational Einstein-Hilbert action ${\mathcal S}_G$ and the matter action ${\mathcal S}_M$, where $\ka = 16\pi G$ and ${\mathcal U}$ is a local potential density for the scalar field. The corresponding equations of motion are
\be
G_{\mu \nu } - \Lambda g_{\mu \nu}= \frac{\kappa}{2} T_{M \mu \nu}, \qquad \nabla_\mu \nabla^\mu \phi = - \frac{\pa {\mathcal U}}{\pa \phi},
\label{20}
\en
where $G_{\mu \nu } = R_{\mu \nu } - \frac{1}{2}Rg_{\mu \nu} $ is the Einstein tensor and
\be
T_{M \mu \nu} =  \frac{2}{\sqrt{-g}} \frac{\de {\mathcal S}_M}{\de g^{\mu \nu}}  = \pa_\mu\phi \pa_\nu\phi -   g_{\mu \nu}\left( \frac{1}{2} g^{\al \beta} \pa_\al\phi \pa_\beta\phi - {\mathcal U}(\phi) \right)   
\label{emtensor}
\en
is the energy-momentum tensor of the matter field.

The action can also be expressed in terms of phase space variables. That is done in the ADM formalism \cite{arnowitt62,bojowald11,carlip19}, resulting in a geometrodynamics. It assumes that space-time can be foliated in terms of space-like hypersurfaces such that the space-time manifold is diffeomorphic to ${\mathbb R} \times \Sigma$, with $\Sigma$ a 3-surface. Coordinates $x^\mu=(t,{\bf x})$ can be chosen such that the time coordinate $t$ labels the leaves of the foliation and ${\bf x}$ are the coordinates on $\Sigma$. In terms of these coordinates, the metric and its inverse are written as
\be
g_{\mu \nu}=
\begin{pmatrix}
N^2 - N_i N^i & -N_i \\
-N_i & - h_{ij}
\end{pmatrix} \,,
\qquad
g^{\mu \nu}=
\begin{pmatrix}
\frac{1}{N^2} & \frac{-N^i}{N^2} \\
\frac{-N^i}{N^2} &   \frac{N^iN^j}{N^2}- h^{ij}
\end{pmatrix} \,,
\label{cqg2}
\en
where $N$ and $N_i$ are respectively the lapse and the shift vector, and $h_{ij}$ is the  Riemannian metric on $\Sigma$. Spatial indices are raised and lowered by this spatial metric. The unit normal vector field to the space-like hypersurfaces is
\be
n^\mu = (1/N, -N^i/N)^T. \label{normal}
\en

The canonically conjugate momenta are 
\be
\pi_\phi = \frac{\pa {\mathcal L}_{\textrm{cl}} }{ \pa {\dot \phi}} =  \frac{\sqrt{h}}{N}\left(\dot \phi - N^i \pa_i \phi \right)
\en
and
\be
\pi^{ij} = \frac{\pa {\mathcal L}_{\textrm{cl}} }{ \pa {\dot h}_{ij}} = - \frac{1}{\kappa} \sqrt{h} (K^{ij} - K h^{ij}),
\en
with 
\be
K_{ij} = \frac{1}{2N}({ D _i N_j + D _j N_i - \dot{h}}_{ij}) ,\qquad K = K_{ij}h^{ij},
\en
the extrinsic curvature. (The conjugate momentum is a tensor density of weight $-1$). In terms of these variables the action \eqref{10} reads
\be
{\mathcal S}_{\textrm{cl}} = \int dt \int_\Sigma d^3x \left( {\dot h}_{ij} \pi^{ij} + \dot \phi \pi_\phi \right)  - \int dt H,
\label{classicalaction}
\en
where $H$ is the Hamiltonian, given by
\be
H = \int_\Sigma d^3 x \left(N  {\mathcal H} + N^i  {\mathcal H}_i  \right),
\en
with
\be
{\mathcal H}  =   \ka G_{ijkl} \pi^{ij} \pi^{kl} + \frac{1}{2} \frac{1}{\sqrt h} \pi^2_\phi + {\mathcal V}(h,\phi), \qquad {\mathcal H}_j  =  -2 h_{jk} D_i \pi^{ik} + \pi_\phi \pa_i \phi,\label{clham}
\en
where $h$ is the determinant of $h_{ij}$, $G_{ijkl} = (h_{ik} h_{jl} + h_{il} h_{jk} -  h_{ij} h_{kl})/2\sqrt{h}$ is the DeWitt metric, with inverse $G^{ijkl} = \sqrt{h}(h^{ik} h^{jl} + h^{il} h^{jk} - 2 h^{ij} h^{kl})/2$ and determinant $\det(G_{ijkl})=-2h$, $D_i$ is the covariant derivative corresponding to the metric $h_{ij}$, and ${\mathcal V}$ is the potential density
\be
{\mathcal V}(h,\phi) = \frac{\sqrt{h}} {\ka}( 2\Lambda - R^{(3)}) +  \sqrt{h} \left( \frac{1}{2} h^{ij} \pa_i\phi \pa_j \phi + {\mathcal U} \right) .
\label{cqg5}
\en
We define
\be
V(t) = \int_\Sigma d^3 x  N({\bf x},t){\mathcal V}(h({\bf x},t),\phi({\bf x},t)).
\label{D3}
\en
In terms of these variables, the equations of motion are{\footnote{Equation \eqref{c11} can even be worked out further, but that is not necessary for our purposes (for the full expression see e.g.\ \cite{wald84,bojowald11}).}}

\be
{\dot{h}}_{ij} = 2\ka NG_{ijkl}\pi^{kl} + D _i N_j + D _j N_i,
\label{c10}
\en
\be
{\dot{\pi}}^{ij} = - N \ka \frac{\pa G_{mnkl}}{\pa h_{ij}} \pi^{mn} \pi^{kl} - \frac{\de V}{ \de h_{ij}} - \frac{\de }{ \de h_{ij}} \int_\Sigma d^3y \left(-2 N_k D_l \pi^{kl} \right) + \frac{N}{2} \frac{ h^{ij}}{\sqrt{h}} \pi^2_\phi  .
\label{c11}
\en
\be
\dot \phi =  \frac{N}{\sqrt h} \pi_\phi + N^i \pa_i \phi, \qquad {\dot \pi}_\phi  = - \frac{\delta V}{\delta \phi} + N^i \pa_i \pi_\phi,
\label{c12}
\en
\be
{\mathcal H} =0, \qquad  {\mathcal H}_i  = 0  .
\label{c13}
\en
The last two equations are the Hamiltonian and diffeomorphism constraint.

\section{Bohmian quantum gravity}\label{bqm}
\subsection{Bohmian geometrodynamics }
Canonical quantization  of the theory leads to the Wheeler-DeWitt theory, where the state is given by a wave functional $\Psi(h,\phi)$ of the 3-metric $h_{ij}({\bf x})$ and the scalar field $\phi({\bf x})$, which satisfies the operator constraints 
\be
{\widehat {\mathcal H}} \Psi = 0,
\label{q1}
\en
\be
{\widehat {\mathcal H}}_i \Psi = 0,
\label{q2}
\en
with{\footnote{There is the usual operator ordering ambiguity in turning classical constraints to operator constraints. We have followed here a particular ordering which corresponds to the Laplace-Beltrami ordering with respect to the DeWitt metric at each point $x$ \cite{kuchar73,christodoulakis86}, with corresponding Bohmian theory given in \cite{horiguchi94,shojai98}. Another ordering is the one of DeWitt \cite{dewitt67a}, which correponds to $ -\kappa G_{ijkl}\frac{\delta^2}{\delta h_{ij} \delta h_{kl}}$ for the kinetic term in \eqref{q3}, with corresponding Bohmian theory given in for example \cite{pinto-neto05a,pinto-neto19}. These choices of orderings do not affect the form of the guidance equations, which are given by \eqref{bcqg1} and \eqref{dotphi} in both cases. The only difference will be in the form of the gravitation quantum potential. Rather than the expression \eqref{D1}, the DeWitt ordering implies that ${\mathcal Q}_{ G} = -\ka G_{ijkl} \frac{1}{|\Psi|}  \frac{\delta^2 |\Psi|}{\delta h_{ij}\delta h_{kl}}$. For the rest, quantities like the energy-momentum tensor like \eqref{q20} remain unchanged.}}
\be
{\widehat {\mathcal H}}   =  - \ka {\sqrt h}\frac{\delta}{\delta h_{ij}} \left( \frac{1}{\sqrt h} G_{ijkl}\frac{\delta}{\delta h_{kl}}\right) - \frac{1}{2}\frac{1}{\sqrt h}\frac{\delta^2}{\delta \phi^2} + {\mathcal V}(h,\phi), 
\label{q3}
\en
\be
{\widehat {\mathcal H}}_i  =  -2 h_{ik}D_j\frac{\delta }{\delta h_{jk}} + \frac{1}{2} \left( \pa_i \phi \frac{\de }{\de \phi} + \frac{\de }{\de \phi} \pa_i \phi\right).
\en
Expressing $\Psi(h,\phi)=|\Psi(h,\phi)|\ee^{\ii S(h,\phi)}$ in polar form, the guidance equation for the {\it actual} 3-metric $h_{ij}$ and the {\it actual} scalar field $\phi$ (playing now the role of the {\it actual} position $x$ above) are \cite{horiguchi94,shtanov96,goldstein04,pinto-neto05a,pinto-neto19}:
\be
{\dot{h}}_{ij} = 2\ka NG_{ijkl}\frac{\delta S}{\delta h_{kl}} + D _i N_j + D _j N_i,
\label{bcqg1}
\en
\be
\dot \phi =   \frac{N}{\sqrt h} \frac{\delta S}{\delta \phi} + N^i \pa_i \phi.
\label{dotphi}
\en

The DeWitt metric $G_{ijkl}$ plays a quite analogous role to the Riemannian metric $g_{ij}$ in non-relativistic Bohmian mechanics. In particular, just like in the non-relativistic guidance equation \eqref{1b}, it is used in \eqref{bcqg1} to define a velocity field for the 3-metric. 

We conclude that \eqref{q1}-\eqref{dotphi} define a Bohmian quantum theory of gravitation. Given an initial 3-metric, a lapse function and a shift function, the dynamics determines a 4-metric, defined by \eqref{cqg2}. Different choices of shift function will lead to the same 4-geometry, i.e., to the same metric up to space-time diffeomorphisms. However, different lapse functions generically will yield different 4-geometries (unless they merely differ by a factor which depends only on time). Therefore a particular choice should be made for the lapse function. In other words, the dynamics depends on the choice of foliation.

Non-relativistic Bohmian mechanics reproduces the predictions of orthodox quantum theory. In the case of quantum gravity, it is unclear what exactly the predictions of the orthodox theory are \cite{isham92,kuchar92}. On the other hand, Bohmian quantum gravity allows for precise predictions, e.g.\ concerning the possible evolutions of the universe. Statistical predictions derive from the typicality measure with density $|\Psi|^2$, which is preserved by the Bohmian dynamics as a result of the continuity equation (which follows from \eqref{q1} and \eqref{q2}):
\begin{multline}
\int_\Sigma d^3x \frac{\delta}{\delta h_{ij}} \left[ \left(2\ka NG_{ijkl}\frac{\delta S}{\delta h_{kl}} + D _i N_j + D _j N_i\right)|\Psi|^2 \right] \\
+ \int_\Sigma d^3x \frac{\de}{\de \phi} \left[\left( \frac{N}{\sqrt h} \frac{\delta S}{\delta \phi} + N^i \pa_i \phi \right)|\Psi|^2 \right] =0 .
\end{multline}
The fact that this measure is not normalizable should be dealt in the way explained in a similar context in \cite{duerr19}.

\subsection{Bohmian geometrodynamics in Hamiltonian form and action\label{sec3}}
To derive the quantum Einstein equations, we first recast the above Bohmian equations in the second-order Hamiltonian form in analogy with the equations \eqref{c10}-\eqref{c13}. From this we can formulate an action principle for the dynamics.

Defining the ``canonical momenta''
\be
\pi^{ij}=\frac{\delta S}{\delta h_{ij}}, \qquad \pi_\phi = \frac{\de S }{\de \phi},
\label{q15}
\en
the guidance equations (cf.\ \eqref{bcqg1} and \eqref{dotphi}) have the classical form
\be
{\dot{h}}_{ij} = 2\ka NG_{ijkl}\pi^{kl} + D _i N_j + D _j N_i, \qquad \dot \phi =  \frac{N}{\sqrt h} \pi_\phi + N^i \pa_i \phi.
\label{h1}
\en

Furthermore, plugging the polar form of $\Psi$ into the constraint \eqref{q1}, it follows that
\be
 \ka G_{ijkl}\frac{\delta S}{\delta h_{ij}} \frac{\delta S}{ \delta h_{kl}} + \frac{1}{2}\frac{1}{\sqrt h}\left(\frac{\delta S}{\delta \phi}\right)^2 + {\mathcal V} + {\mathcal Q}=0,
\label{q10}
\en
where ${\mathcal Q}$ is the quantum potential density, given by
\be\label{qptot}
{\mathcal Q}(h,\phi) = {\mathcal Q}_{ G}(h,\phi) + {\mathcal Q}_{M}(h,\phi),
\en
with
\be
{\mathcal Q}_{ G} = -\ka \frac{\sqrt h}{|\Psi|}\frac{\delta}{\delta h_{ij}} \left( \frac{1}{{\sqrt h}} G_{ijkl}\frac{\delta |\Psi|}{\delta h_{kl}}  \right), \label{D1}
\en 
\be
{\mathcal Q}_{M} =- \frac{1}{2{\sqrt h}|\Psi|} \frac{\delta^2 |\Psi|}{\delta \phi^2}\label{D2}\,.
\en
Unlike ${\mathcal V}$, ${\mathcal Q}$ is not a local functional of the fields.

The constraint \eqref{q2} implies that 
\be
-2 D_j\frac{\delta S}{\delta h_{ji}} + \pa^i \phi \frac{\de S }{\de \phi} =0.
\label{q11}
\en
Hence, using \eqref{q15}, we have that
\be
{\mathcal H} + {\mathcal Q} = 0, \qquad {\mathcal H}^i = 0.
\label{h2}
\en
These are the classical constraint equations \eqref{c13} modified by ${\mathcal V} \to {\mathcal V} + {\mathcal Q}$. 

Finally, using \eqref{q10} and \eqref{q11}, and defining
\be
Q(t) = \int_\Sigma d^3 x  N({\bf x},t){\mathcal Q}(h({\bf x},t),\phi({\bf x},t)),
\en
it can straightforwardly (but somewhat tediously) be shown that from the Bohmian dynamics (cf.\ \eqref{bcqg1} and \eqref{dotphi}),  the classical equations \eqref{c11}, \eqref{c12} follow --- merely changed by an additional term which arises from the replacement ${\mathcal V} \to {\mathcal V} + {\mathcal Q}$, as expected:
\be
{\dot{\pi}}^{ij} = - N \ka \frac{\pa G_{mnkl}}{\pa h_{ij}} \pi^{mn} \pi^{kl} - \frac{\de }{ \de h_{ij}}(V+Q) - \frac{\de }{ \de h_{ij}} \int_\Sigma d^3y \left(-2 N_k D_l \pi^{kl} \right) + \frac{N}{2} \frac{ h^{ij}}{\sqrt{h}} \pi^2_\phi  .
\label{h3}
\en
\be
\dot \phi =  \frac{N}{\sqrt h} \pi_\phi + N^i \pa_i \phi, \qquad {\dot \pi}_\phi  = - \frac{\delta }{\delta \phi} (V+Q) + N^i \pa_i \pi_\phi.
\label{h4}
\en

Summarizing, we get the classical equations \eqref{c10}-\eqref{c13} but with an additional potential ${\mathcal Q}$, which amounts to replacing ${\mathcal V}$ by $ {\mathcal V} + {\mathcal Q}$. Hence, with the replacement $H \to H + \int_\Sigma d^3 x N{\mathcal Q}$ in the classical action \eqref{classicalaction}, we can derive the Hamiltonian form of the Bohmian dynamics from the action
\be
{\mathcal S} = {\mathcal S}_{\textrm{cl}} + {\mathcal S}_{Q},
\label{baction}
\en
with 
\be
{\mathcal S}_Q = - \int dt \int_\Sigma d^3x N{\mathcal Q} =  - \int dt \int_\Sigma d^3x N{\sqrt h} \frac{{\mathcal Q}}{{\sqrt h}} =  - \int dt \int_\Sigma d^3x {\sqrt{ -g}} \frac{{\mathcal Q}}{{\sqrt h}},
\en
and observing the constraint \eqref{q15}. Actually it is sufficient to assume that \eqref{q15} holds on a certain leaf of the foliation \cite{pinto-neto99}. Then it will hold on all leaves of the foliation.
 
 Note that the wave equations \eqref{q1} and \eqref{q2} remain unchanged. It is only in the equations for the Bohmian variables $h_{i j}, \phi$ or better their ``canonical momenta'' $\pi^{ij},\pi_{\phi}$  that the quantum potential appears.

\section{Quantum Einstein equations from the action}\label{quantumeinstein1}
The most direct way to find the quantum Einstein equations is by considering the definition of the Einstein tensor $G_{\mu \nu}$ (as a function of the metric) and to see what the Bohmian equations of motion entail for the  total energy-momentum tensor. We will do that in the next section. Here, we will first follow the route outlined in the introduction for nonrelativistic Bohmian mechanics by resorting to an action principle (as in \cite{shojai98}).

The Hamiltonian form of the Bohmian dynamics derives from the action \eqref{baction}, which was expressed in terms of phase space variables. Viewing it as a functional of the fields $g_{\mu \nu}$ and $\phi$ (which is immediate since the quantum potential is just a functional of the fields and not of the momenta), variation with respect to $g_{\mu \nu}$ gives the quantum Einstein equations:
\be
G_{\mu \nu } - \Lambda g_{\mu \nu}= \frac{\kappa}{2} \left( T_{M \mu \nu} + T_{Q \mu \nu}\right),
\label{qee}
\en
where $T_{M \mu \nu}$ is given in \eqref{emtensor}  and the quantum term is{\footnote{The expression for $T_{Q}^{\mu\nu}$ is different from the one given in \cite{shojai98,shojai02,shojai04} where at least the first term of \eqref{qem} seems missing.}}
\begin{align}
T_{Q}^{\mu \nu}(x) &= -\frac{2}{\sqrt{-g(x)}} \frac{\de {\mathcal S}_Q}{\de g_{\mu \nu}(x)} \nonumber\\
&= g^{\mu \nu} (x) \frac{{\mathcal Q}(x)}{\sqrt{h(x)}} + \frac{2}{\sqrt{-g(x)}} \int dt' d^3y {\sqrt{ -g({\bf y},t')}} \frac{\de}{\de g_{\mu \nu}(x)} \frac{{\mathcal Q}({\bf y},t')}{{\sqrt{h({\bf y},t')}}},
\label{q20}
\end{align}
with ${\mathcal Q}({\bf x},t) = {\mathcal Q}(h({\bf x},t),\phi({\bf x},t))$ given by \eqref{qptot}-\eqref{D2}. More explicitly, the components are{\footnote{Note that the indices of the energy-momentum tensor are raised and lowered using the space-time metric $g_{\mu \nu}$ and {\it not} by the spatial metric $h_{ij}$.\label{raising}}} 
\be T_{Q}^{00}({\bf x},t) = \frac{1}{N({\bf x},t)^2 } \frac{{\mathcal Q}({\bf x},t)}{\sqrt{h({\bf x},t)}}    , \quad  T_{Q}^{0i}({\bf x},t) = T_{Q}^{i0}({\bf x},t) =  - \frac{N^i({\bf x},t)}{N ({\bf x},t)^2} \frac{{\mathcal Q}({\bf x},t)}{\sqrt{h({\bf x},t)}}, \en 
\begin{align} T_{Q}^{ij}({\bf x},t) &= \left(\frac{N^i({\bf x},t)N^j({\bf x},t)}{N({\bf x},t)^2} - h^{ij}({\bf x},t)\right)  \frac{{\mathcal Q}({\bf x},t)}{\sqrt{h({\bf x},t)}} \nonumber\\
& \qquad - \frac{2}{N({\bf x},t)\sqrt{h({\bf x},t)}}  \int_\Sigma d^3y  N({\bf y},t)\sqrt{h({\bf y},t)} \frac{\de }{\de h_{ij}({\bf x})} \left( \frac{{\mathcal Q}({\bf y},t)}{\sqrt{h({\bf y},t)}} \right)  \nonumber\\
& = \frac{N^i({\bf x},t)N^j({\bf x},t)}{N({\bf x},t)^2} \frac{{\mathcal Q}({\bf x},t)}{\sqrt{h({\bf x},t)}} -  \frac{2}{N({\bf x},t)\sqrt{h({\bf x},t)}}\frac{\de }{\de h_{ij}({\bf x})}\int_\Sigma d^3y N({\bf y},t){\mathcal Q}({\bf y},t)
.
\label{qem}
\end{align}
Using the normal vector field $n^\mu$ defined in \eqref{normal}, we can also write this tensor as
\be
T_{Q}^{\mu \nu}(x) = n^\mu(x) n^\nu(x) \frac{{\mathcal Q}(x)}{\sqrt{h(x)}} - \frac{2}{\sqrt{-g(x)}} \de^\mu_i \de^\nu_j \int_\Sigma d^3y N({\bf y},t)\frac{\de {\mathcal Q}({\bf y},t)}{\de h_{ij}({\bf x})}  ,
\label{q21}
\en
or, defining ${\widetilde {\mathcal Q}} = {\mathcal Q}/\sqrt{h}$, 
\be
T_{Q}^{\mu \nu}(x) = g^{\mu \nu}(x) {\widetilde {\mathcal Q}}(x) - \frac{2}{\sqrt{-g(x)}} \de^\mu_i \de^\nu_j \int_\Sigma d^3y {\sqrt{-g({\bf y},t)}}   \frac{\de {\widetilde {\mathcal Q}}({\bf y},t)}{\de h_{ij}({\bf x})}  .
\label{q21.b}
\en

The scalar field satisfies
\be
\nabla_\mu \nabla^\mu \phi({\bf x},t) = - \frac{\pa {\mathcal U} ( \phi({\bf x},t))}{\pa \phi}- \frac{1}{\sqrt{-g({\bf x},t)}} \frac{\de Q(t)}{\de \phi({\bf x})} 
\en
or 
\be
\nabla_\mu \nabla^\mu \phi = - \frac{\pa {\mathcal U} }{\pa \phi}- \frac{1}{\sqrt{-g}} \frac{\de Q}{\de \phi} .
\en

While the quantum Einstein equations looks covariant, they are not. As mentioned before, the dynamics depends on the choice of foliation, i.e., the choice of lapse function.

\section{Direct route to the quantum Einstein equations}\label{quantumeinstein2}
The quantum Einstein equations can also be derived more directly, although a bit more tediously. This alternative route is useful to present, since it could be used for a Bohmian theory for which there is no action governing the second-order dynamics (which might for example be the case when matter is described by the Dirac theory). The Einstein tensor is a particular function of the metric and its space-time derivatives up to second order. The expression for the Bohmian energy-momentum tensor can then be found by substitution of the time derivatives of the metric using the Bohmian dynamics \eqref{bcqg1} and \eqref{dotphi}.

We start with splitting up the equations of motions in terms of the classical gravity part, classical matter part and quantum part. The constraint equations \eqref{h2} can respectively be written as 
\be
{\mathcal H}_G + {\mathcal H}_M + {\mathcal H}_Q = 0 , \qquad {\mathcal H}_{Gi} + {\mathcal H}_{Mi} = 0,
\label{d1}
\en
with (recall for that also \eqref{clham}, \eqref{cqg5} and \eqref{D3})
\be
{\mathcal H}_G = \ka G_{ijkl} \pi^{ij} \pi^{kl} + \frac{\sqrt{h}} {\ka}( 2\Lambda - R^{(3)})    , 
\en
\be
{\mathcal H}_M =  \frac{1}{2} \frac{1}{\sqrt h} \pi^2_\phi +  \sqrt{h} \left( \frac{1}{2} h^{ij} \pa_i\phi \pa_j \phi + {\mathcal U} \right) , \qquad {\mathcal H}_Q = {\mathcal Q} ,
\en
\be
{\mathcal H}_{Gi} = -2 h_{ik} D_j \pi^{jk}, \qquad {\mathcal H}_{Mi}  = \pi_\phi \pa_i \phi.
\en

In addition, the equation \eqref{h3} for $\dot \pi^{ij}$ is given by 
\be
{\mathcal F}^{ij}_G + {\mathcal F}^{ij}_M + {\mathcal F}^{ij}_Q = 0, 
\label{d2}
\en
where
\be
{\mathcal F}^{ij}_G = {\dot{\pi}}^{ij} + N \ka \frac{\pa G_{mnkl}}{\pa h_{ij}} \pi^{mn} \pi^{kl} + \frac{\de }{ \de h_{ij}} \int_\Sigma d^3y \left[\frac{N\sqrt{h}} {\ka}( 2\Lambda - R^{(3)})  -2 N_k D_l \pi^{kl} \right] ,
\en
\be
{\mathcal F}^{ij}_M = - \frac{N}{2} \frac{ h^{ij}}{\sqrt{h}} \pi^2_\phi  + \frac{\de }{ \de h_{ij}}\int_\Sigma d^3y N\sqrt{h} \left( \frac{1}{2} h^{kl} \pa_k\phi \pa_l \phi + {\mathcal U} \right)  ,
\en
\be
{\mathcal F}^{ij}_Q =  \frac{\de Q}{ \de h_{ij}} .
\en

The left-hand side of the Einstein field equations is given by \cite{bertschinger02}{\footnote{The comment in footnote \ref{raising} applies again.}
\begin{align}
G^{00} - \Lambda g^{00} &= -\frac{\ka}{2} \frac{ {\mathcal H}_G}{N^2{\sqrt{h}}},\nonumber\\ 
G^{0i} - \Lambda g^{0i}&= \frac{\ka}{2}  \frac{N{\mathcal H}^i_G + N^i{\mathcal H}_G } {N^2 {\sqrt{h}} } ,\nonumber\\ 
G^{ij} - \Lambda g^{ij}&= -\frac{\ka}{2} \frac{ N^i N^j {\mathcal H}_G}{N^2{\sqrt{h}}} + \ka \frac{1}{N {\sqrt{h}} }{\mathcal F}^{ij}_G,
\label{d3}
\end{align}
where the momenta $\pi^{ij}$ and their derivatives should be expressed in terms of the metric $h_{ij}$ and its derivatives using \eqref{h1}. Namely, the form of the tensor $G^{\mu\nu}-\Lambda g^{\mu\nu}$ is that of the free classical theory, for which the same relation between $\pi^{ij}$ and their derivatives with the metric $h_{ij}$ and its derivatives hold.

For free classical gravity, the equations of motion are ${\mathcal H}_G = 0$, ${\mathcal H}^i_G=0$ and ${\mathcal F}^{ij}_G=0$, which yields $G^{\mu\nu}-\Lambda g^{\mu\nu}=0$. In the Bohmian case, we have instead \eqref{d1} and \eqref{d2}, and from \eqref{d3} we can immediately read of that \eqref{qee} must hold, with
\begin{align}
T^{00}_M &=  \frac{ {\mathcal H}_M}{N^2{\sqrt{h}}},\nonumber\\ 
T^{0i}_M &= -  \frac{N{\mathcal H}^i_M + N^i{\mathcal H}_M } {N^2 {\sqrt{h}} } ,\nonumber\\ 
T^{ij}_M &= \frac{ N^i N^j {\mathcal H}_M}{N^2{\sqrt{h}}} - 2\frac{1}{N {\sqrt{h}} }{\mathcal F}^{ij}_M,
\label{d4}
\end{align}
which is the energy-momentum tensor \eqref{emtensor} in terms of phase space variables, and
\begin{align}
T^{00}_Q &=  \frac{ {\mathcal H}_Q}{N^2{\sqrt{h}}},\nonumber\\ 
T^{0i}_Q &= -  \frac{N^i {\mathcal H}_Q } {N^2 {\sqrt{h}} } ,\nonumber\\ 
T^{ij}_Q &= \frac{ N^i N^j {\mathcal H}_Q}{N^2{\sqrt{h}}} - 2\frac{1}{N {\sqrt{h}} }{\mathcal F}^{ij}_Q,
\label{d5}
\end{align}
which is again the expression \eqref{qem} for the quantum part of the energy-momentum tensor.

\section{Mini-superspace}\label{Mini-superspace}
We can also give the quantum Einstein equations in the case of a homogeneous and isotropic mini-superspace. The mini-superspace model is a simplified version of quantum gravity which arises by using the usual quantization techniques to the symmetry reduced classical theory. So this theory is not derived from the full Wheeler-DeWitt theory, but serves as a toy model for quantum gravity. 

Classically, the homogeneity and isotropy imply that 
\be
\phi = \phi(t), \qquad ds^2 = g_{\mu \nu} (t)dx^\mu dx^\nu = N^2(t) dt^2 - a^2(t) d \Omega^2_k, 
\en
with $a$ the scale factor and $d\Omega^2_k$ the spatial line element with constant spatial curvature $k$. The energy-momentum tensor is that of a perfect fluid:
\be
T^{\mu \nu} = (\rho + p ) U^\mu U^\nu - p g^{\mu \nu}
\en
where $U^\mu$ is the four-velocity which is $U^\mu = \delta^\mu_0/N$ in the comoving frame, and  
\be
\rho = \frac{\dot \phi^2}{2N^2} + {\mathcal U}(\phi) , \qquad p = \frac{\dot \phi^2}{2N^2} - {\mathcal U}(\phi).
\label{m10}
\en
are the density and pressure. The Einstein equations can be brought into the familiar form of the Friedmann equations
\be
\frac{1 }{ N^2}  \frac{\dot a^2 }{ a^2} = \frac{8\pi G}{3} \rho - \frac{k}{a^2} + \frac{\Lambda}{3} ,
\label{m11}
\en
\be
\frac{1}{aN} \frac{d}{dt}\left( \frac{\dot a}{N}\right) = - \frac{4\pi G}{3} (\rho + 3 p) + \frac{\Lambda}{3}.
\label{m12}
\en
The scalar field equation reads
\be
\frac{1}{N}\frac{d}{dt}\left( \frac{a^3 \dot \phi}{N}\right) +  a^3 \pa_\phi {\mathcal U} = 0.
\en
Using \eqref{m11} and the definitions \eqref{m10}, this equation can be written as 
\be
\dot \rho = -3\frac{\dot a}{a}(\rho + p).
\label{m13}
\en
This equation can also be derived from \eqref{m11} and \eqref{m12}, and actually expresses the conservation of the energy-momentum tensor.

The Wheeler-DeWitt equation in the mini-superspace theory is (see \cite{pinto-neto19}):
\be
\left[- \frac{1}{2a^3} \pa^2_\phi + \frac{{\bar \ka}^2 }{2 a^3} (a \pa_a)^2 + a^3 \left( {\mathcal U}(\phi) - \frac{k}{2{\bar \ka}^2a^2} + \frac{\Lambda}{6{\bar \ka}^2} \right) \right] \psi(a,\phi) = 0,
\en
with ${\bar \ka}^2 = 4\pi G/3$, and writing $\psi = |\psi|\ee^{\ii S}$ the guidance equations are
\be
\dot \phi = \frac{N}{ a^3} \pa_\phi S ,\qquad \dot a = -{\bar \ka}^2\frac{N}{ a} \pa_a S.
\en
It is straightforward to check that these equations imply the classical equations \eqref{m11}, \eqref{m12} and \eqref{m13}, with the replacement $\rho \to \rho + \rho_Q$ and $p \to p + p_Q$ (see \cite{pinto-neto19} for some details), where the quantum density 
\be
\rho_Q = \frac{1}{a^3}{\mathcal Q} = - \frac{1}{2a^6} \frac{\pa^2_\phi |\psi|}{|\psi|} + \frac{{\bar \ka}^2}{2a^6} \frac{(a\pa_a )^2 |\psi|}{|\psi|}
\en
depends on the quantum potential density ${\mathcal Q}$, and the quantum pressure is 
\be
p_Q = - \frac{1}{3a^2} \pa_a {\mathcal Q} .
\en
So the quantum contribution to the energy-momentum is 
\be
T^{\mu \nu}_Q= (\rho_Q + p_Q ) U^\mu U^\nu - p_Q g^{\mu \nu}.
\en

\section{Conservation of energy}\label{energy}
The quantum Einstein equations immediately imply that the energy-momentum tensor is covariantly conserved: Since $\nabla_\mu G^{\mu \nu} \equiv 0$, we must have that $\nabla_\mu (T^{\mu \nu} + T^{\mu \nu}_Q)= 0$ along a Bohmian path. So the classical conservation equation holds. This should be contrasted to the situation in other cases \cite{maudlin19}, such as non-relativistic Bohmian mechanics or the Bohmian theories in an external {\em classical} space-time (e.g.\ Minkowski space-time), where the classical conservation equations of respectively energy conservation and covariant conservation of the energy-momentum tensor do not hold. While the violation of the classical conservation laws in the latter cases in no way impairs those theories, it is interesting that in Bohmian quantum gravity one has as much as one can hope for classically concerning the conservation of the energy-momentum tensor.

Consider first non-relativistic Bohmian mechanics. The total energy $E = m v^2/2 + V + Q$, defined in analogy with Newtonian mechanics,  is generically not conserved along a trajectory. (If $Q$ is does not explicitly depend on time, like for an energy eigenstate, then the total energy {\em is} conserved.) So in this case the classical conservation law does not hold. 

Consider now a Bohmian scalar field theory in an external classical space-time \cite{struyve15} (with Minkowski space-time as a special case). The Schr\"odinger equation for the wave functional $\Psi(\phi,t)$ reads
\be
\ii \frac{\pa \Psi}{\pa t} = \int_\Sigma d^3 x \left[ N \sqrt{ h} \left(-\frac{1}{2h} \frac{\de^2}{\de \phi^2}  + \frac{1}{2} h^{ij} \pa_i \phi \pa_j \phi + \mathcal{U} \right)  -   \frac{\ii}{2} N^i \left(\pa_i \phi \frac{\de}{\de \phi}  + \frac{\de}{\de \phi} \pa_i \phi \right)  \right] \Psi \,, 
\label{e1}
\en
and the guidance equation is
\be
{\dot \phi} = \frac{N}{\sqrt{h}} \frac{\de S}{\de \phi} + N^i \pa_i \phi .
\label{e2}
\en
As a natural choice for the energy-momentum-tensor we can take the one that appears in \eqref{qee}, but now computed for a wave function that does not depend on the metric.\footnote{In \cite{struyve15} the tensor was derived  by applying the strategy of sections \ref{sec3} and \ref{quantumeinstein1} to \eqref{e2}} Thus 
\be
T^{\mu \nu} =  T^{\mu \nu}_M + T_Q^{\mu \nu} \,, 
\label{e3}
\en
with $T^{\mu \nu}_M$ the classical energy-momentum tensor \eqref{emtensor} and $T_Q^{\mu \nu}$ quantum potential tensor 
\be \label{qphdep}
T_Q^{\mu \nu} =  \left( n^\mu n^\nu + \de^\mu_i\de^\nu_j h^{ij}  \right) \frac{{\mathcal{Q}_M}}{\sqrt h} , \quad {\mathcal Q}_{M} =- \frac{1}{2{\sqrt h}|\Psi|} \frac{\delta^2 |\Psi|}{\delta \phi^2}.
\en
(Observe that in the computing the latter from \eqref{q21}, the quantum potential depends on the metric through the factor $\sqrt{h}$.) However, the tensor $T^{\mu \nu}$ is not covariantly conserved. Namely,
\be
\nabla_\mu T^{\mu \nu} =  -  \frac{1}{\sqrt{ -g}} \pa^\nu \phi  \frac{\de }{\de \phi} \int_\Sigma d^3x N{\mathcal{Q}_M}  + \nabla_\mu T_Q^{\mu \nu}  \,,
\en
which is generically different from zero.{\footnote{Even if we took the Bohmian energy-momentum tensor to be given by just $T^{\mu \nu}_M$, i.e., without the addition of $ T_Q^{\mu \nu}$ in \eqref{e3}, then still the tensor would not be covariantly  conserved.} Also in the special case of Minkowski space-time, still the energy-momentum tensor is generically not conserved. (In the special case of an energy-eigenstate, the total energy $\int d^3 x T^{00}$ is conserved.)

In the search for a semi-classical theory for gravity, where gravity is treated classically, it is a natural idea is to use the Bohmian energy-momentum tensor $T^{\mu \nu}$ source the gravitational field in the classical Einstein field equations. Contrary to the theory defined by \eqref{e1} and \eqref{e2}, there would then be a back-reaction from quantum matter onto gravity. Such a semi-classical theory could be seen as an approximation to quantum gravity or possibly even be conceived of as a ``fundamental'' theory. However, the fact that the Bohmian energy-momentum tensor \eqref{e3} is not conserved poses a consistency problem (which is absent in full quantum gravity). As explained in detail in \cite{struyve15}, this problem can hopefully be overcome by suitably dealing with the spatial diffeomorphism invariance.

\section{Conclusion}\label{con}
Bohmian quantum gravity entails deviations from the classical dynamics. One way to capture these deviations is by looking at the deviations from the classical Einstein field equations. These deviations appear as an extra contribution to the energy-momentum tensor that is of quantum mechanical origin, i.e., it depends on the quantum potential. These deviations can lead to effects like space-time singularity resolution \cite{pinto-neto12b,falciano15,pinto-neto19} or perhaps an effective cosmological constant \cite{demaerel20}. 

We have only performed the analysis for the case matter is described by a scalar field and a Bohmian field ontology. Instead of a field ontology, one could also consider a Bohmian formulation in terms of a particle ontology, i.e., by introducing actual material point-particles. This would be naturally done in the context of the spin-1/2 Dirac theory. Alternatively, one could also consider again the spin-0 scalar field theory. The advantage would be the greater simplicity, but the disadvantage is the familiar problematic behaviour already in Minkowski space-time \cite{holland93b}, such as space-like particle velocities. Rather than considering quantum field theory (i.e., the second quantized level) one could consider just a fixed number of particles described by the Klein-Gordon theory. The Wheeler-DeWitt theory for this case was presented in \cite{pavsic11}.

\section{Acknowledgments}
It is a pleasure to thank Thibaut Demaerel, Christian Maes and Kasper Meerts for useful discussions. While at the LMU Munich, WS was supported by the Deutsche Forschungsgemeinschaft. Currently, WS is supported by the Research Foundation Flanders (Fonds Wetenschappelijk Onderzoek, FWO), Grant No. G066918N.

\end{document}